\documentclass[nofootinbib,superscriptaddress,twocolumn]{revtex4-1}

\usepackage{amsmath,amsfonts,amssymb}
\usepackage{wrapfig}
\usepackage{graphicx}
\usepackage{bbm}
\usepackage{color}
\usepackage{float}
\usepackage{subfigure}
\usepackage{textcomp}
\usepackage[british]{babel}
\usepackage{pifont}

\usepackage[dvipsnames]{xcolor}
\usepackage[colorlinks=true,linkcolor=blue,citecolor=blue]{hyperref}

\usepackage{wasysym}
\usepackage{harmony}
\newcommand{\ket}[1]{\left|#1\right\rangle }

\usepackage{bbm}

\usepackage{xspace} 
\newcommand*{\melvin}{{\small M}{\scriptsize ELVIN}\xspace}

\begin{document}

\title{Active learning machine learns to create new quantum experiments}

\author{Alexey A. Melnikov}
\thanks{These authors contributed equally to this work}
\affiliation{Institute for Theoretical Physics, University of Innsbruck, Technikerstra{\ss }e 21a, 6020 Innsbruck, Austria}

\author{Hendrik Poulsen Nautrup}
\thanks{These authors contributed equally to this work}
\affiliation{Institute for Theoretical Physics, University of Innsbruck, Technikerstra{\ss }e 21a, 6020 Innsbruck, Austria}

\author{Mario Krenn}
\affiliation{Vienna Center for Quantum Science and Technology, Faculty of Physics, University of Vienna, Boltzmanngasse 5, 1090 Vienna, Austria}
\affiliation{Institute for Quantum Optics and Quantum Information, Austrian Academy of Sciences, Boltzmanngasse 3, 1090 Vienna, Austria}

\author{Vedran Dunjko}
\affiliation{Institute for Theoretical Physics, University of Innsbruck, Technikerstra{\ss }e 21a, 6020 Innsbruck, Austria}
\affiliation{Max-Planck-Institute for Quantum Optics, Hans-Kopfermann-Str. 1, 85748 Garching, Germany}

\author{Markus Tiersch}
\affiliation{Institute for Theoretical Physics, University of Innsbruck, Technikerstra{\ss }e 21a, 6020 Innsbruck, Austria}

\author{Anton Zeilinger}
\affiliation{Vienna Center for Quantum Science and Technology, Faculty of Physics, University of Vienna, Boltzmanngasse 5, 1090 Vienna, Austria}
\affiliation{Institute for Quantum Optics and Quantum Information, Austrian Academy of Sciences, Boltzmanngasse 3, 1090 Vienna, Austria}

\author{Hans J. Briegel}
\affiliation{Institute for Theoretical Physics, University of Innsbruck, Technikerstra{\ss }e 21a, 6020 Innsbruck, Austria}
\affiliation{Department of Philosophy, University of Konstanz, Fach 17, 78457 Konstanz, Germany}


\begin{abstract}
How useful can machine learning be in a quantum laboratory? Here we raise the question of the potential of intelligent machines in the context of scientific research. A major motivation for the present work is the unknown reachability of various entanglement classes in quantum experiments. We investigate this question by using the projective simulation model, a physics-oriented approach to artificial intelligence. In our approach, the projective simulation system is challenged to design complex photonic quantum experiments that produce high-dimensional entangled multiphoton states, which are of high interest in modern quantum experiments. The artificial intelligence system learns to create a variety of entangled states, and improves the efficiency of their realization. In the process, the system autonomously (re)discovers experimental techniques which are only now becoming standard in modern quantum optical experiments -- a trait which was not explicitly demanded from the system but emerged through the process of learning. Such features highlight the possibility that machines could have a significantly more creative role in future research.
\end{abstract}

\maketitle

\section{Introduction}
Automated procedures are indispensable in modern science. Computers rapidly perform large calculations, help us visualize results, and specialized robots perform preparatory laboratory work to staggering precision. But what is the true limit of the utility of machines for science? To what extent can a machine help us understand experimental results? Could \emph{it} -- the machine -- discover new useful experimental tools? The impressive recent results in the field of machine learning, from reliably analysing photographs~\cite{NIPS2012_4824} to beating the world champion in the game of Go~\cite{Silver-Hassabis2016}, support optimism in this regard. Researchers from various research fields now employ machine learning algorithms~\cite{Jordan255}, and the success of machine learning applied to physics~\cite{CarrasquillaMelko2017, VanNieuwenburgLiuHuber2017, SchmidtLipson2009, carleo2016solving} in particular is already noteworthy. Machine learning has claimed its place as the new data analysis tool in the physicists toolbox~\cite{zdeborova2017machine}. However, the true limit of machines lies beyond data analysis, in the domain of the broader theme of artificial intelligence (AI), which is much less explored in this context. In an AI picture, we consider devices (generally called \emph{intelligent agents}) that interact with an environment (the laboratory) and learn from previous experience~\cite{RusselNorvig2003}. To broach the broad questions raised above, in this work we address the specific yet central question of whether an intelligent machine can propose novel and elucidating quantum experiments. We answer in the affirmative in the context of photonic quantum experiments, although our techniques are more generally applicable. We design a learning agent, which interacts with (the simulations of) optical tables, and learns how to generate novel and interesting experiments. More concretely, we phrase the task of the development of new optical experiments in a reinforcement learning (RL) framework~\cite{barto1998reinforcement}, vital in modern AI~\cite{2015_QL_Nature,Silver-Hassabis2016}.

The usefulness of automated designs of quantum experiments has been shown in Ref.~\cite{PhysRevLett.Mario}. There, the algorithm \melvin starts from a toolbox of experimentally available optical elements to randomly create simulations of setups. Those setups are reported if they result in high-dimensional multipartite entangled states. The algorithm uses handcrafted rules to simplify setups and is capable of learning by extending its toolbox with previously successful experimental setups. Several of these experimental proposals have been implemented successfully in the lab~\cite{malik2016multi, SchledererKrennFicklerMalikZeilinger2016cyclic,babazadeh2017high,erhard2017experimental}, and have led to the discovery of new quantum techniques~\cite{KrennHochrainerLahiriZeilinger2016, krenn2017quantum}.

\begin{figure*}[ht!]
	\centering
	\includegraphics[width=1\textwidth]{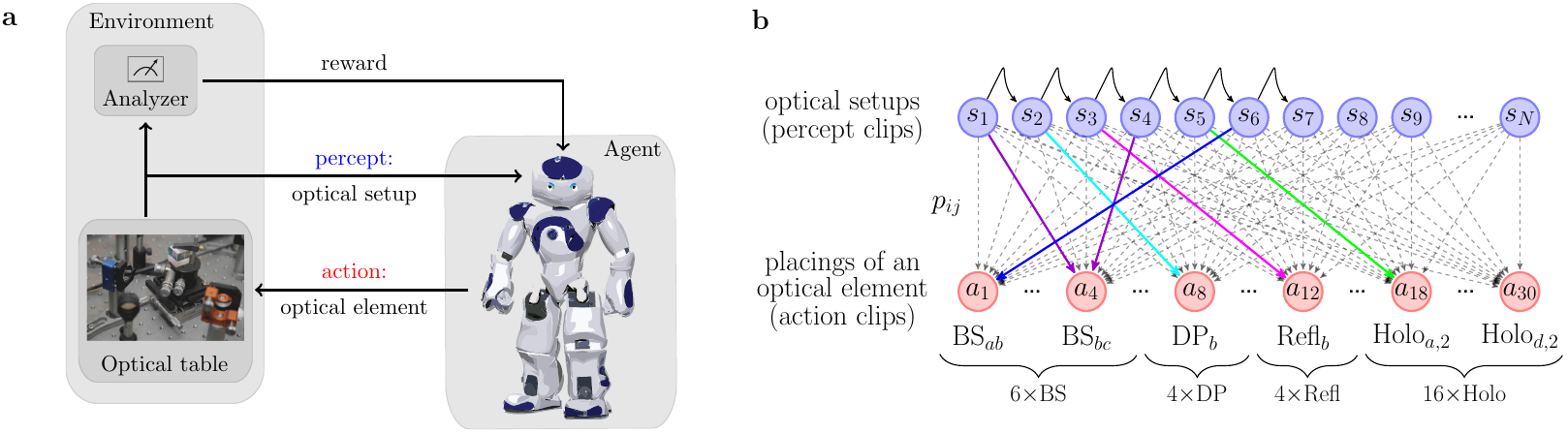}
	\caption{The learning agent. (a) An agent is always situated in an environment~\cite{RusselNorvig2003}. Through sensors it perceives optical setups and with actuators it can place optical elements in an experiment. Note that, in this paper, the interaction between the agent and environment was entirely simulated on a classical computer.	[One could imagine that in the future, a real robot builds up the experiment designed by the computer.] On the side, an analyzer evaluates a proposed experiment corresponding to the current optical setup and gives rewards according to a specified task. (b) The memory network that represents the internal structure of the PS agent. Dashed arrows indicate possible transitions from percept clips (blue circles) to action clips (red circles). Solid, colored arrows depict a scenario where a sequence of actions leads to the experiment $\{\mathrm{BS}_{bc}, \mathrm{DP}_b, \mathrm{Refl}_b, \mathrm{BS}_{bc}, \mathrm{Refl}_{b}, \mathrm{Holo}_{a,2}\}$. Arrows between percepts correspond to deterministic transitions from one experiment to another after placement of an optical element.}
	\label{fig:Agent}
\end{figure*}

Inspired by the success of the \melvin algorithm in the context of finding specific optical setups, here we investigate the broader potential of learning machines and artificial intelligence in designing quantum experiments~\cite{2012_Briegel}. Specifically, we are interested in their potential to contribute to novel research, beyond rapidly identifying solutions to fully specified problems. To investigate this question, we employ a more general model of a learning agent, formulated within the projective simulation (PS) framework for artificial intelligence~\cite{2012_Briegel}, which we apply to the concrete test-bed of Ref.~\cite{PhysRevLett.Mario}.
In the process of generating specified optical experiments, the learning agent builds up a memory network of correlations between different optical components -- a feature it later exploits when asked to generate targeted experiments efficiently. In the process of learning, it also develops notions (technically speaking, composite clips~\cite{2012_Briegel}), for components that ``work well'' in combination. In effect, the learning agent autonomously discovers sub-setups (or gadgets) which are useful outside of the given task of searching for particular high-dimensional multiphoton entanglement\footnote{The discovery of such devices is in fact a byproduct stemming from the more involved structure of the learning model.}.

We concentrate on the investigation of multipartite entanglement in high dimensions~\cite{PhysRevA.89.012105,HuberDeVicenteJulio2013, HuberPerarnau-LlobetDeVicenteJulio2013} and give two examples where learning agents can help. The understanding of multipartite entanglement remains one of the outstanding challenges of quantum information science, not only because of the fundamental conceptual role of entanglement in quantum mechanics, but also because of the many applications of entangled states in quantum communication and computation~\cite{RevModPhys.81.865,RevModPhys.80.517,RevModPhys.84.777,PhysRevA.69.062311, PhysRevLett.87.117901,scarani2009security}. As the first example, the agent is tasked to find the simplest setup that produces a quantum state with a desired set of properties. Such efficient setups are important as they can be robustly implemented in the lab. The second task is to generate as many experiments, which create states with such particular properties, as possible. Having many different setups available is important for understanding the structure of the state space reachable in experiments and for exploration of different possibilities that are accessible in experiments. In both tasks the desired property is chosen to be a certain type of entanglement in the generated states. More precisely we target high-dimensional ($d>2$) many-particle ($n>2$) entangled states. The orbital angular momentum (OAM) of photons~\cite{allen1992orbital, Mair2001, molina2007twisted, krenn2017orbital} can be used for investigations into high-dimensional~\cite{vaziri2002experimental,dada2011experimental, agnew2011tomography, krenn2014generation, zhang2016engineering} or multiphoton entanglement~\cite{hiesmayr2016observation, wang2015quantum} and, since recently, also both simultaneously~\cite{malik2016multi}. OAM setups have been used as test-beds for other new techniques for quantum optics experiments~\cite{PhysRevLett.Mario, KrennHochrainerLahiriZeilinger2016}, and they are the system of choice in this work as well.

\section{Results}
Our learning setup can be put in terms of a scheme visualized in Fig.~\ref{fig:Agent}(a)\footnote{Part of this figure is adapted from Ref.~\cite{web:robot}. Image of the optical table is a part of the $(3,3,3)$-experiment (see main text) in Ref.~\cite{erhard2017experimental}, by Manuel Erhard (University of Vienna).}. The agent (our learning algorithm) interacts with a virtual environment -- the simulation of an optical table. The agent has access to a set of optical elements (its  toolbox), with which it generates experiments: it sequentially places the chosen elements on the (simulated) table; following the placement of an element, the quantum state generated by the corresponding setup i.e. configuration of optical elements is analyzed. Depending on the state, and the chosen task, the agent either receives a reward or not. The agent then observes the current table configuration and places another element; thereafter, the procedure is iterated. We are interested in finite experiments, so the maximal number of optical elements in one experiment is limited. This goes along with practical restrictions: due to accumulation of imperfections, e.g., through misalignment and interferometric instability, for longer experiments we expect a decreasing overall fidelity of the resulting state or signal. The experiment ends (successfully) when the reward is given or (unsuccessfully) when the maximal number of elements on the table is reached without obtaining the reward. Over time, the agent learns which element to place next, given a table.

\begin{figure*}[ht!]
	\centering
	\includegraphics[width=1\textwidth]{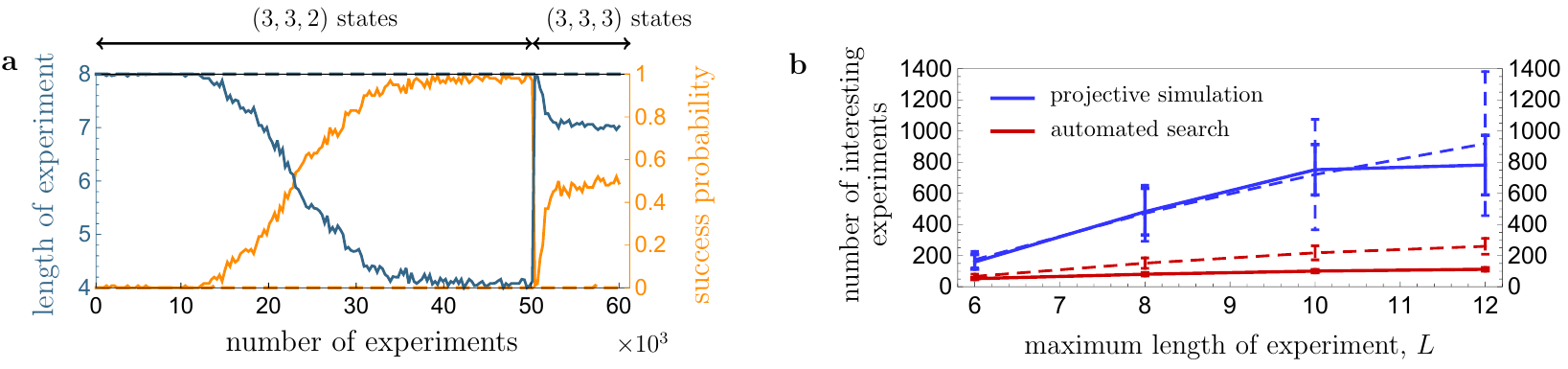}
	\caption{Results of learning new experiments. (a) Average length of experiment and success probability in each of the $6\times 10^4$ experiments. The maximal length of an experiment is $L=8$. During the first $5\times 10^4$ experiments an agent is rewarded for obtaining a $(3,3,2)$ state, during the last $10^4$ experiments the same agent is rewarded when finding a $(3,3,3)$ state. The average success probability shows how probable it is for the PS agent to find a rewarded state in a given experiment. Solid/dashed lines show simulations of the PS agent that learns how to generate a $(3,3,3)$ state from the beginning with/without prior learning of setups that produce a $(3,3,2)$ state. (b) Average number of interesting experiments obtained after placing $1.2\times 10^5$ optical elements. Data points are shown for $L=6, 8, 10$ and $12$. Dashed/solid blue and red lines correspond to PS with/without action composition and automated random search~\cite{PhysRevLett.Mario} with/without action composition, respectively (see main text). Vertical bars indicate the mean squared deviation. (a), (b) All data points are obtained by averaging over $100$ agents. Parameters of the PS agents are specified in \textit{Supporting Information}.
	}
	\label{fig:Results}
\end{figure*}

Initially, we fix the analyzer in Fig.~\ref{fig:Agent}(a) to reward the successful generation of a state out of a certain class of high-dimensional multipartite entangled states. Conceptually, this choice of the reward function explicitly specifies our criterion for an experimental setup (and the states that are created) to be \textit{interesting}, but we will come back to a more general scenario momentarily. We characterize interesting states by a Schmidt-Rank Vector (SRV)~\cite{HuberDeVicenteJulio2013, HuberPerarnau-LlobetDeVicenteJulio2013} -- the numerical vector containing the rank of the reduced density matrix of each of the subsystems -- together with the requirement that the states are maximally entangled in the OAM basis~\cite{PhysRevLett.Mario, malik2016multi}.

The initial state of the simulated experiment is generated by a double spontaneous parametric down-conversion (SPDC) process in two nonlinear crystals. Similar to Refs.~\cite{PhysRevLett.Mario, malik2016multi}, we ignore higher-order terms with OAM $|m|>1$ from down-conversion as their amplitudes are significantly smaller than the low-order terms. Neglecting these higher-order terms in the down-conversion, the initial state $\ket{\psi(0)}$ can be written as a tensor product of two pairs of OAM-entangled photons~\cite{malik2016multi},
\begin{equation}
	\ket{\psi(0)} = \frac{1}{3}\left(\sum_{m=-1}^{1} \ket{m}_a\ket{-m}_b\right) \otimes \left(\sum_{m=-1}^{1} \ket{m}_c\ket{-m}_d\right),
\label{eq:initial}
\end{equation}\\
where the indices $a, b, c$ and $d$ specify four arms in the optical setup. The toolbox contains a basic set of elements~\cite{PhysRevLett.Mario} including beam splitters (BS), mirrors (Refl), shift-parametrized holograms (Holo) and Dove prisms (DP). Taking into account that each distinct element can be placed in any one (or two in the case of BS) of the four arms $a,b,c,d$ we allow in total $30$ different choices of elements. Since none of the optical elements in our toolbox creates entanglement in OAM degrees of freedom, we use a measurement in arm $a$ and post-selection to ``trigger'' a tripartite state in the other three arms.

Our learning agent is based on the projective simulation~\cite{2012_Briegel} model for AI. PS is a physics-motivated framework which can be used to construct RL agents. PS was shown to perform well in standard RL problems~\cite{2013_Mautner_PSII,Grid-world-PS,bjerland2015projective,melnikov2017projective}, in advanced robotics applications~\cite{simon2016}, and it is also amenable for quantum enhancements~\cite{paparo2014quantum,dunjko2015quantum,friis2015coherent}. The main component of the PS agent is its memory network (shown in Fig.~\ref{fig:Agent}(b)) comprising units of episodic memory called \emph{clips}. Here, clips include remembered \emph{percepts} (in our case, the observed optical tables) and \emph{actions} (corresponding to the placing of an optical element). Each percept (clip) $s_i$, $i\in[1, \dots, N]$ is connected to every action (clip) $a_j$, $j\in[1, \dots, 30]$ via a weighted directed edge $(i,j)$, which represents the possibility of taking an action $a_j$ in a situation $s_i$ with probability $p_{i,j}$, see Fig.~\ref{fig:Agent}(b). The process of learning is manifested in the creation of new clips, and in the adjustment of the probabilities. Intuitively, the probabilities of percept-action transitions which eventually lead to a reward will be enhanced, leading to a higher likelihood of rewarding behaviour in the future (see \textit{Supporting Information} for details).

\subsection{Designing short experiments}
As mentioned, a maximum number of elements placed in one experiment is introduced to limit the influence of experimental imperfections. For the same reason, it is valuable to identify the shortest experiments that produce high-dimensional tripartite entanglement characterized by a desired SRV. In order to relate our work to existing experiments~\cite{malik2016multi}, we task the agent with designing a setup that creates a state with SRV $(3,3,2)$. To further emphasize the benefits of learning, we then investigate whether the agent can use its knowledge, attained through the learning setting described previously, to discover a more complex quantum state. Thus, the task of finding a state with SRV $(3,3,2)$ is followed by the task of finding a state with SRV $(3,3,3)$, after $5\times10^4$ simulated experiments. As always, a reward is issued whenever a target state is found, and the table is reset. Fig.~\ref{fig:Results}(a) shows the success probability throughout the learning process: the PS agent first learns to construct a $(3,3,2)$ state, and then, with probability $0.5$, very quickly (compared to the by itself simpler task of finding a (3,3,2) state) learns to design a setup that corresponds to a $(3,3,3)$ state. Our results suggest that either the knowledge of constructing a $(3,3,2)$ state is highly beneficial in the second phase, or that the more complicated $(3,3,3)$ state is easier to generate. To resolve this dichotomy, we simulated a learning process where the PS agent is required to construct a $(3,3,3)$ state within $6\times 10^4$ experiments, without having learned to build a $(3,3,2)$ state during the first $5\times 10^4$ experiments. The results of the simulations are shown in Fig.~\ref{fig:Results}(a) as dashed lines (lower and upper edge of the frame). It is apparent that the agent without previous training on $(3,3,2)$ states does not show any significant progress in constructing a $(3,3,3)$ experiment. Furthermore, the PS agent constantly and autonomously improves by constructing shorter and shorter experiments (cf. Fig.~S1(a) in \textit{Supporting Information}). By the end of the first phase, PS almost always constructs experiments of length~$4$ -- the shortest length of an experiment producing a state from the $(3,3,2)$ SRV class. During the $(3,3,3)$ learning phase, the PS agent produces a $(3,3,3)$ state of the shortest length in half of the cases. Experimental setups useful for the second phase are almost exclusively so-called parity sorters (see Fig.~\ref{usefulExp}). Other setups that are not equivalent to parity sorters seem not to be beneficial in finding a $(3,3,3)$ state. As we will show later, the PS agent tends to use parity sorters more frequently while exploring the space of different SRV classes. This is particularly surprising since the parity sorter itself was originally designed for a different task.

\subsection{Designing new experiments}
The connection between high-dimensional entangled states, and the structure of optical tables which generate them is not well understood~\cite{PhysRevLett.Mario}. Having a database of such experiments would allow us to deepen our understanding of the structure of the set of entangled states that can be accessed by optical tables. In particular, such a database could then be further analyzed to identify useful sub-setups or \textit{gadgets}~\cite{PhysRevLett.Mario} -- certain few-element combinations that appear frequently in larger setups -- which are useful for generating complex states. With our second task, we have challenged the agent to generate such a database by finding high-dimensional entangled states. As a bonus, we found that the agent does the post-processing above for us implicitly, and in run-time. The outcome of this post-processing is encoded in the structure of the memory network of the PS agent. Specifically, the sub-setups which were particularly useful in solving this task are clearly visible, or embodied, in the agent's memory.

To find as many different high-dimensional three-photon entangled states as possible, we reward the agent for every new implementation of an interesting experiment. To avoid trivial extensions of such implementations, a reward is given only if the obtained SRV was not reached before within the same experiment. Fig.~\ref{fig:Results}(b) displays the total number of new, interesting experiments designed by the basic PS agent (solid blue) and the PS agent with action \emph{composition}~\cite{2012_Briegel} (dashed blue). Action composition allows the agent to construct new composite actions from useful optical setups (i.e. placing multiple elements in a fixed configuration), thereby autonomously enhancing the toolbox (see \textit{Supporting Information} for details). It is a central ingredient for an AI to exhibit even a primitive notion of creativity~\cite{Briegel2012} and was also used in Ref.~\cite{PhysRevLett.Mario} to augment automated random search. For comparison, we provide the total number of interesting experiments obtained by automated random search with and without action composition (solid and dashed red curves). As we will see later, action composition will allow for additional insight into the agent's behavior and helps provide useful information about quantum optical setups in general. We found that the PS model discovers significantly more interesting experiments than both automated random search and automated random search with action composition, see Fig.~\ref{fig:Results}(b).

\begin{figure*}[ht!]
	\centering
	\includegraphics[width=18cm]{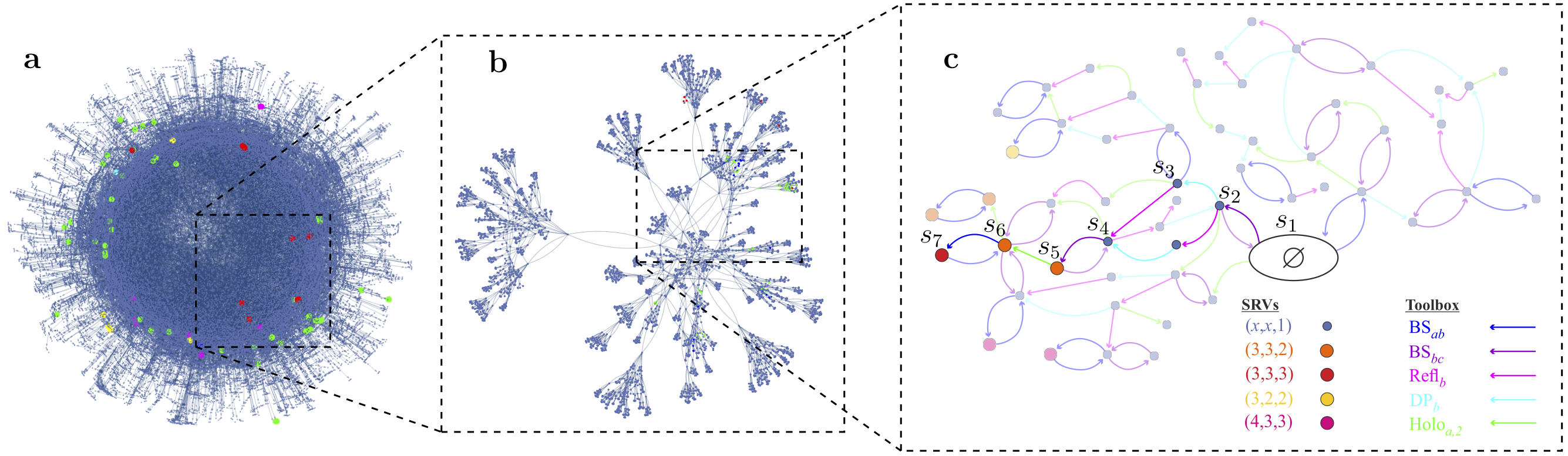}
	\caption{Exploration space of optical setups. Different setups are represented by vertices with colors specifying an associated SRV (biseparable states are depicted in blue). Arrows represent the placing of optical elements.
		(a) A randomly generated space of optical setups. Here we allow up to $6$ elements on the optical table and a standard toolbox of $30$ elements. Large, colored vertices represent interesting experiments. If two nodes share a color, they can generate a state with the same SRV. Running for $1.6\times 10^4$ experiments, the graph that is shown here has $45605$ nodes, out of which $67$ represent interesting setups.
		(b) A part of the graph (a), which demonstrates the nontrivial structure of the network of optical setups.
		(c) A detailed view of one part of the bigger network. The depicted colored maze represents an analogy between the task of finding the shortest implementation of an experiment and the task of navigating in a maze~\cite{barto1998reinforcement, Grid-world-PS, mirowski2016learning, hierarchical2016maze}.
		Arrows of different colors represent distinct optical elements that are placed in the experiment. The initial state is represented by an empty table $\varnothing$. The shortest path to a setup that produces a state with SRV $(3,3,2)$ and $(3,3,3)$ is highlighted. Labels along this path coincide with the labels of the percept clips in Fig.~\ref{fig:Agent}(b).
	}
	\label{fig:beautifulMaze}
\end{figure*}

\subsection{Ingredients for successful learning}
In general, successful learning relies on a structure hidden in the task environment (or dataset). The results presented thus far show that PS is highly successful in the task of designing new interesting experiments, and here we elucidate why this should be the case. The following analysis also sheds light on other settings where we can be confident that RL techniques can be applied as well.

First, the space of optical setups can be illustrated using a graph as given in Fig.~\ref{fig:beautifulMaze}(c), where the building of an optical experiment corresponds to a walk on the directed graph. Note that optical setups that create a certain state are not unique: two or more different setups can generate the same quantum state. Due to this fact, this graph does not have a tree structure but rather resembles a maze. Navigating in a maze, in turn, constitutes one of the classic textbook RL problems~\cite{barto1998reinforcement, Grid-world-PS, mirowski2016learning, hierarchical2016maze}. Second, our empirical analysis suggests that experiments generating high-dimensional multipartite entanglement tend to have some structural similarities~\cite{PhysRevLett.Mario} (see Fig.~\ref{fig:beautifulMaze}(a)-(b) which partially displays the exploration space). The figure shows regions where the density of interesting experiments (large colored nodes) is high and others where it is low -- interesting experiments seem to be clustered (cf. Fig.~S2). In turn, RL is particularly useful when one needs to handle situations which are similar to those previously encountered -- once one maze (optical experiment) is learned, similar mazes (experiments) are tackled more easily, as we have seen before. In other words, whenever the experimental task has a maze-type underlying structure, which is often the case, PS can likely help -- and critically, without having any {\em a priori} information about the structure itself~\cite{Grid-world-PS, 2016_Makmal_ML}. In fact, PS gathers information about the underlying structure throughout the learning process. This information can then be extracted by an external user or potentially be utilized further by the agent itself.

\subsection{The potential of learning from experiments}
Thus far we have established that a machine can indeed design new quantum experiments in the setting where the task is precisely specified (via the rewarding rule). Intuitively, this could be considered the limit of what a machine can do for us, as machines are specified by our programs. However, this falls short from what, for instance, a human researcher can achieve. How could we, even in principle, design a machine to do something (interesting) we have not specified it to do? To develop an intuition for the type of behavior we could hope for, consider, for the moment, what we may expect a human, say a good PhD student would do in situations similar to those studied thus far.

To begin with, a human cannot go through all conceivable optical setups to find those that are interesting. Arguably, she would try to identify prominent sub-setups and techniques that are helpful in the solving of the problem. Furthermore, she would learn that such techniques are probably useful beyond the specified tasks and may provide new insights in other contexts. Could a machine, even in principle, have such insight? Arguably, traces of this can be found already in our, comparatively simple, learning agent.
By analyzing the memory network of the agent (ranking clips according to the sum of the weights of their incident edges), specifically the composite actions it learned to generate, we can extract sub-setups that have been particularly useful in the endeavor of finding many different interesting experiments in Fig.~\ref{fig:Results}(b) for $L=6$.

\begin{figure}[ht!]
	\begin{center}
		\includegraphics[width=8.7cm]{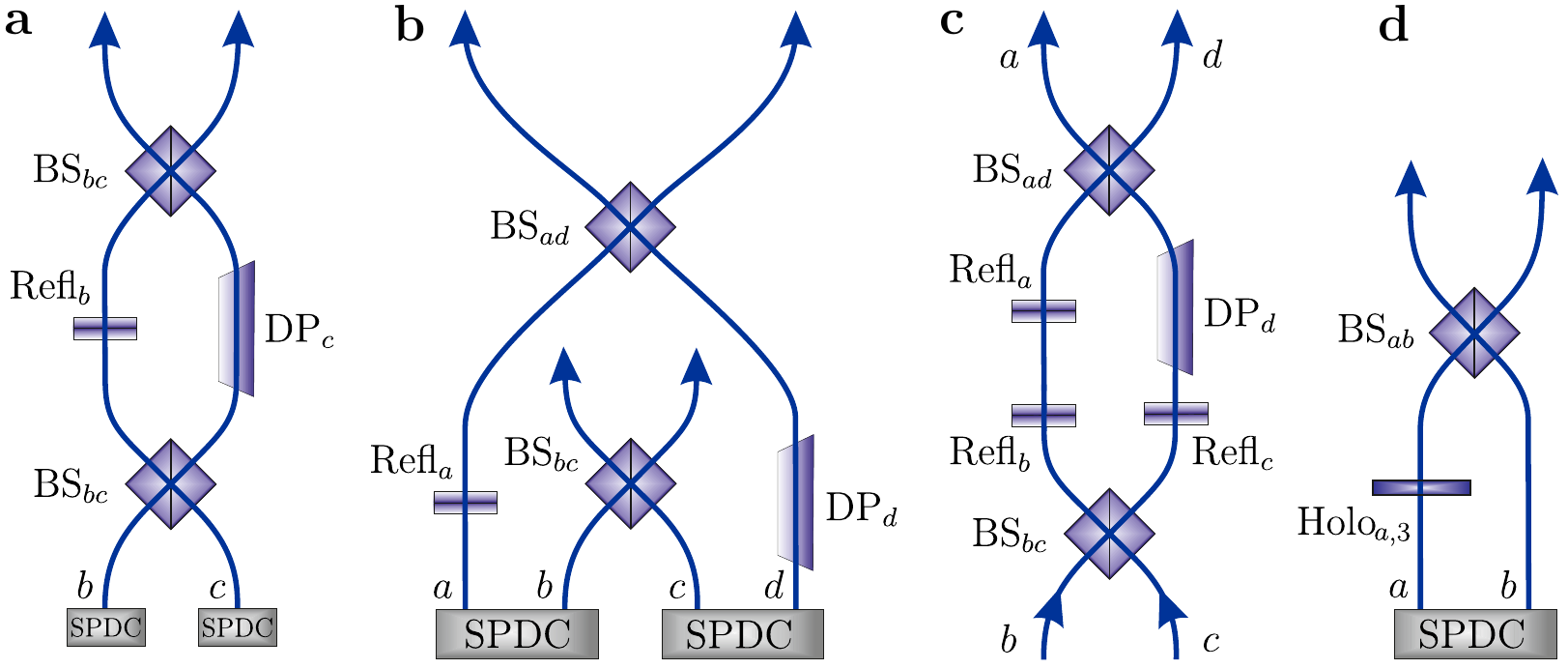}
	\end{center}
	\vspace{-0.6cm}
	\caption{Experimental setups frequently used by the PS agent. (a) Local parity sorter. (b) Nonlocal parity sorter (as discovered by the program). (c) Nonlocal parity sorter in the Klyshko wave front picture~\cite{klyshko1988simple}, in which the paths $a$ and $d$ are identical to the paths $b$ and $c$, respectively. (d) Setup to increase dimensionality of photons. (a-d) In a simulation of 100 agents, the highest weighted sub-setups have been 11 times experiment $a$, 22 times experiment $b$ and 43 times experiment $d$ was part of the highest-weighted subsetup. Only in 24 cases, other sub-setups have been the highest weighted.}
	\label{usefulExp}
\end{figure}

For example, PS composes and then extensively uses a combination corresponding to an optical interferometer as displayed in Fig.~\ref{usefulExp}(a) which is usually used to sort OAM modes with different parities~\cite{Leach-Courtial2002} -- in essence, the agent (re)discovered a parity sorter. This interferometer has already been identified as an essential part of many quantum experiments that create high-dimensional multipartite entanglement~\cite{PhysRevLett.Mario}, especially those involving more than two photons (cf. Fig.~\ref{fig:Results}(a) and Ref.~\cite{malik2016multi}).

One of the PS agent's assignments was to discover as many different interesting experiments as possible. In the process of learning, even if a parity sorter was often (implicitly) rewarded, over time it will no longer be, as in this scenario only novel experiments are rewarded. This, again implicitly, drives the agent to ``invent'' new, different-looking configurations, which are similarly useful.

Indeed, in many instances the most rewarded action is no longer the original parity sorter in form of a Mach-Zehnder interferometer (cf. Fig.~\ref{usefulExp}(a)) but a nonlocal version thereof (see Fig.~\ref{usefulExp}(b)). As it turns out, the two setups are equivalent in the Klyshko wave front picture~\cite{klyshko1988simple,aspden2014experimental}, where the time of two photons in arms $b$ and $c$ in Fig.~\ref{usefulExp}(b) is inverted, and these photons are considered to be reflected at their origin (represented by $\mathrm{Refl}_x$) -- the nonlinear crystals. This transformation results to a reduction of the $4$-photon experiment in Fig.~\ref{usefulExp}(b) to the $2$-photon experiment shown in Fig.~\ref{usefulExp}(c).

Such a nonlocal interferometer has only recently been analysed and motivated the work in Ref.~\cite{KrennHochrainerLahiriZeilinger2016}. Furthermore, by slightly changing the analyzer to only reward interesting experiments that produce states with SRV other than the most common SRV $(3,3,2)$, sub-setups aside from the OAM parity sorter become apparent. For example, the PS discovered and exploited a technique to increase the dimensionality of the OAM of the photons by shifting the OAM mode number in one path and subsequently mixing it with an additional path as displayed in Fig.~\ref{usefulExp}(d). Moreover, one can observe that this technique is frequently combined with a local OAM parity sorter. This setup allows the creation of high-dimensional entangled states beyond the initial state dimension of $3$.

All of the observed setups are, in fact, modern quantum optical gadgets/devices (designed by humans) that either already found applications in state-of-the-art quantum experiments~\cite{malik2016multi} or could be used (as individual tool) in future experiments which create high-dimensional entanglement stating from lower-dimensional entanglement.

\vspace{1cm}
\section{Discussion}
One of the fastest growing trends in recent times is the development of ``smart'' technologies. Such technologies are not only permeating our everyday lives in the forms of ``smart'' phones, ``smart'' watches, and, in some places even ``smart'' self-driving cars, but are expected to induce the next industrial revolution~\cite{Schwab2017}. Hence, it should not come as a surprise when ``smart laboratories'' emerge. Already now, modern labs are to a large extent automated~\cite{King2009}, and are removing the need for human involvement in tedious (or hazardous) tasks.

In this work we broach the question of the potential of automated labs, trying to understand to what extent machines could not only help in research, but perhaps even genuinely perform it. Our approach highlights two aspects of learning machines, both of which will be assets in quantum experiments of the future. First, we have improved upon the original observation that search algorithms can aid in the context of finding special optical set-ups~\cite{PhysRevLett.Mario} by using more sophisticated learning agents. This yields confidence that even more involved techniques from AI research (e.g. generalization~\cite{melnikov2017projective}, meta-learning~\cite{2016_Makmal_ML}, etc, in the context of the PS framework) may yield ever improving methods for the autonomous design of experiments.
In a complementary direction, we have shown that the structure of learning models commonly applied in the context of AI research (even the modest basic PS reinforcement learning machinery augmented with action-clip composition~\cite{2012_Briegel, 2013_Mautner_PSII}), possibly allows machines to tackle problems they were not directly instructed or trained to solve. This supports the expectation that AI methodologies will genuinely contribute to research and, very optimistically, the discovery of new physics.

\section*{ACKNOWLEDGMENTS}
A.A.M., H.P.N., V.D., M.T., and H.J.B. were supported by the Austrian Science Fund (FWF) through Grants No. SFB FoQuS F4012 and DK-ALM: W1259-N27, the Templeton World Charity Foundation through Grant No. TWCF0078/AB46, and by the Ministerium f\"{u}r Wissenschaft, Forschung, und Kunst Baden-W\"{u}rttemberg (AZ: 33-7533.-30-10/41/1).
M.K. and A.Z. were supported by the Austrian Academy of Sciences (\"{O}AW), by the European Research Council (SIQS Grant No. 600645 EU-FP7-ICT), and the Austrian Science Fund (FWF) with SFB FoQuS F40 and FWF project CoQuS No. W1210-N16.

\bibliography{bibliography}

\appendix*
\section*{Supporting Information}
\subsection{Projective simulation}
In the following, we describe the PS model and specify the structure of the learning agent used in this paper. For detailed exposition of the PS model, we refer the reader to Refs.~\cite{2012_Briegel, 2013_Mautner_PSII}. The PS model is based on information processing in a specific memory system, called episodic and compositional memory, which is represented as a weighted network of clips. Clips are the units of episodic memory, which are remembered percepts and actions, or short sequences thereof~\cite{2012_Briegel}. The PS agent perceives a state of an environment, deliberates using its network of clips and outputs an action as shown schematically in Fig.~1. The weighted network of clips is a graph with clips as vertices and directed edges that represent the possible (stochastic) transitions between clips. In this paper, the clips represent the percepts and actions, and directed edges connect each percept to every action, as illustrated in Fig.~1(b). An edge $(i,j)$ connects a percept $s_i, i\in [1, \dots, N(t)]$ to an action $a_j, j \in [1, \dots, K(t)]$, where $N(t)$ and $K(t)$ are the total number of percepts and actions at time step $t$, respectively (each time step corresponds to one interaction cycle shown in Fig.~1(a)). Each edge has a time-dependent weight $h^{(t)}_{ij}$, which defines the probability $p^{(t)}_{ij}$ of a transition from percept $s_i$ to action $a_j$ happening. This transition probability is proportional to the weights, also called $h$-values, and is given by the following expression:
\begin{equation}
	p^{(t)}_{ij} = \frac{h^{(t)}_{ij}}{\sum_{k=1}^{K(t)} h^{(t)}_{ik}}.
\label{policy}
\end{equation}
Initially, all $h$-values are set to $1$, hence the probability of performing action $a_j$ is equal to $p_{ij} = 1/K$, that is, uniform. In other words, the initial agent's behaviour is fully random.

As random behaviour is rarely optimal, changes in the probabilities $p_{ij}$ are neccessary. These changes should result in an improvement of the (average) probability of success, i.e. in the chances of receiving a nonnegative reward $\lambda^{(t)}$. Learning in the PS model is realized by making use of feedback from the environment, to update the weights $h_{ij}$. The matrix with entries $h_{ij}$, the so-called $h$-matrix, is updated after each interaction with an environment according to the following learning rule:
\begin{equation}
	h^{(t+1)} = h^{(t)} - \gamma(h^{(t)}-\mathbbm{1}) + \lambda^{(t)} g^{(t+1)},
\end{equation}
where $\mathbbm{1}$ is an ``all-ones'' matrix of a suitable size and $\gamma$ is the damping parameter of the PS model, responsible for forgetting. The $g$-matrix, or so-called glow matrix, allows to internally redistribute rewards such that decisions that were made in the past are rewarded less than the more recent decisions.
The $g$-matrix is updated in parallel with the $h$-matrix. Initially all $g$-values are set to zero, but if the edge $(i,j)$ was traversed during the last decision-making process, the corresponding glow value $g_{i,j}$ is set to $1$.
In order to determine the extent to which previous actions should be internally rewarded, the $g$-matrix is also updated after each interaction with an environment:
\begin{equation}
	g^{(t+1)} = (1-\eta) g^{(t)},
\end{equation}
where $\eta$ is the so-called glow parameter of the PS model. While finding the optimal values for $\gamma$ and $\eta$ parameters is, in general, a difficult problem, these parameters can be learned by the PS model itself~\cite{2016_Makmal_ML}, however, at the expense of longer learning times. In this paper $\gamma$ and $\eta$ parameters were set to time-independent values without any extensive search for optimal values, which could, in principle, further improve our results.

In addition to the basic PS model described above, two additional mechanisms were used that dynamically alter the structure of the PS network: action composition and clip deletion.

\subsubsection{Action composition}
When action composition~\cite{2012_Briegel} is used, new actions can be created as follows: if a sequence of $l\in(1, \dots, L)$ optical elements (actions) results in a rewarded state, then this sequence is added as a single, composite, action to the agent's toolbox. Having such additional actions available allows the agent to access the rewarded experiments in a single decision step, without reproducing the placement of optical elements one by one. Moreover, composite actions can be placed at any time during an experiment which effectively increases the likelihood of an experiment to contain the composite action as sub-setup. Although the entire sequence of $l$ elements can be represented as a single action, this action still has length $l$. That is, the agent is not allowed to place such an element if there are already $L-l+1$ optical elements on the table.

Because of the growing size of the action space, rewards coming from the analyzer are rescaled proportionally to the size of the toolbox $K(t)$,
\begin{equation}
	 \lambda^{(t)} \rightarrow \lambda \frac{K(t)}{K(0)}
\end{equation}
where $K(0)$ is the size of the initial toolbox. This reward rescaling compensates for an otherwise smaller change in transition probabilities in Eq.~\ref{policy}. Although the action composition leads to only a modest improvement in the assigned tasks (see Fig.~2(b)), it is vital for the creative features exhibited by the agent which we discussed in the Results section.

\begin{figure*}[ht!]
	\centering
	\includegraphics[width=1\textwidth]{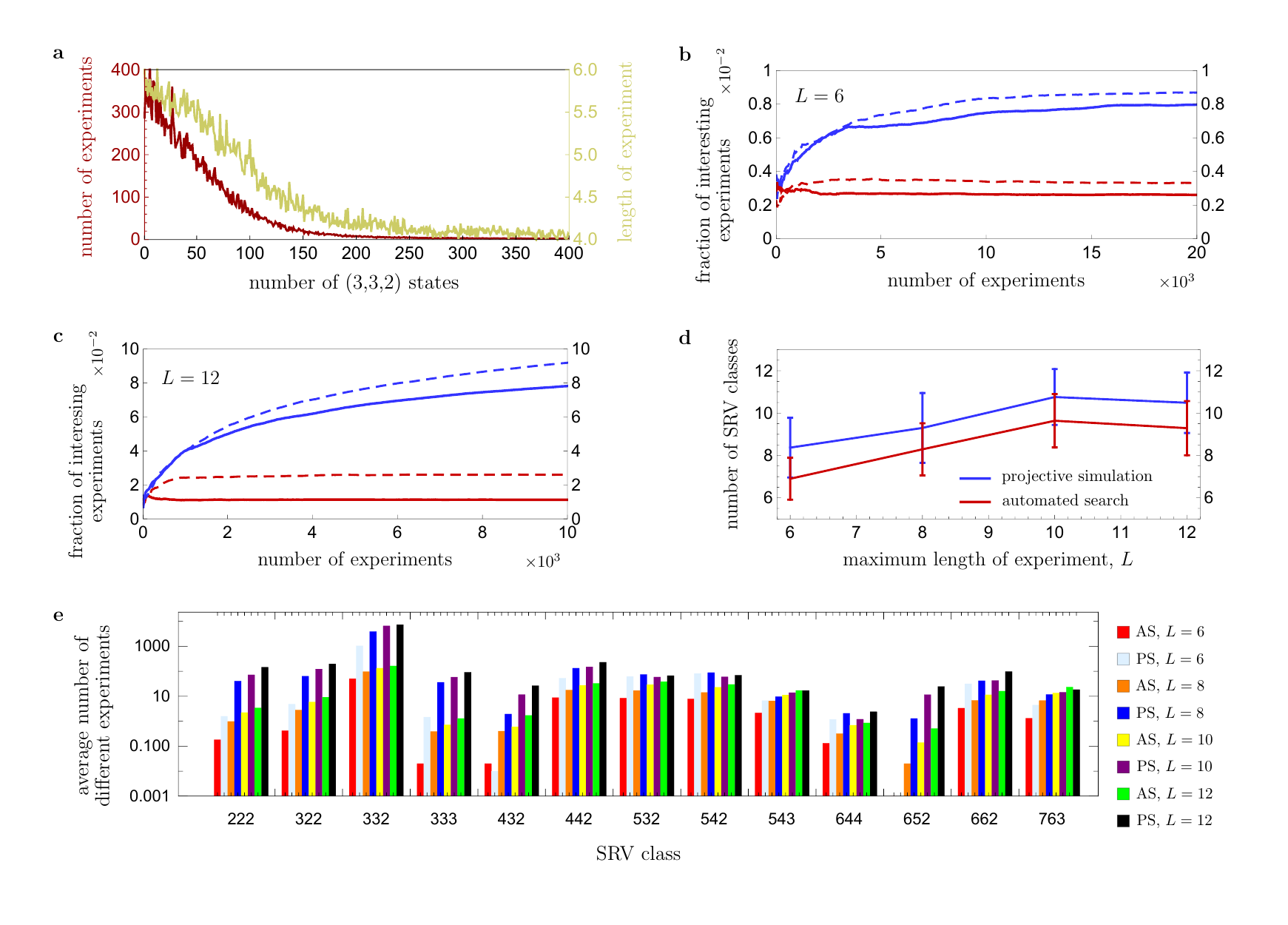}
	\caption{Details of the learning performance. (a) The average number and length of experiments performed before finding a $(3,3,2)$ experiment as a function of number of $(3,3,2)$ states already found. (b), (c) Average fraction of interesting experiments (see main text for definitions) in the cases of $L=6$ (b) and $12$ (c) is shown as a function of the number of experiments. (d) Average number of different SRV classes observed by PS and automated random search as a function of $L$. (e) Average number of different experiments that produce states from different SRV classes are shown for PS and automated random search (AS). (a)-(e) All curves and data points are obtained by averaging over $100$ agents.}
	\label{fig:Methods}
\end{figure*}

\subsubsection{Clip deletion}
Clip deletion enables the agent to handle a large percept and action space by deleting clips that are used only infrequently. Since the underlying mechanisms for percept and action creation are different, the deletion mechanisms are also treated differently.

Let us start by describing the percept deletion mechanism. By keeping all the percepts, the number of possible table configurations scales in highest order as $\mathcal{O}(K^L)$. For example, the simplest situation that we study ($K=30$ and $L=6$) comprises more than $0.5$ billion table configurations. In order to circumvent the storing of all configurations of the percept space, we delete percepts that correspond to unsuccessful experiments since they will not be of any relevance in the future.
If, after placing $L$ elements on the optical table, no positive reinforcement was obtained, all clips and edges that were created during this experiment are deleted. This deletion mechanism allows us to maintain a compact size for the PS network with no more than $10^4$ percept clips on average in the most complicated scenario that we considered ($L=12$).

Now, let us consider action deletion. Due to action composition the number of actions $K$ grows with the number of rewarded experiments. Without deleting action clips, the number of actions can easily go beyond $K=100$, making basic elements less likely to be chosen. In order to keep the balance between the \textit{exploitation} of new successful sequences and the \textit{exploration} of new actions, we assign to each action clip $a_j$, with $j>K(0)$, a probability $p^\mathrm{del}_{a_j}(t)$ to be deleted together with all corresponding edges. These probabilities are given by
\begin{equation}
	p^\mathrm{del}_{a_j}(t) = \left(\frac{N(t)}{\sum_{k=1}^{N(t)} h_{kj}(t)}\right)^{N(t)}.\label{eq:actiondel}
\end{equation}
Action deletion is triggered just after a reward is obtained and deactivated again until a new reward is encountered.
The intuition behind Eq.~\ref{eq:actiondel} is the following. If the in-going edges of action $a_j$ have, to a large extent, comparatively small weights (i.e. either the action has not received many rewards or the $h$-values have been damped significantly since the last rewards were received), then the probability $p^\mathrm{del}_{a_j}(t)$ can be approximated as
\begin{equation}
	p^\mathrm{del}_{a_j}(t) = \left(\frac{N(t)+N_R(t)}{N(t)}\right)^{-N(t)} \approx 1-N_R(t)
\label{ProbDelet}
\end{equation}
with $N_R(t)=\sum_{k=1}^{N(t)} h_{kj}(t)-N(t)<1$. Initially, after the new action has been created, $N_R(t)$ tends to be smaller than one since the sum of in-bound $h$-values is approximately equal to the number of ingoing edges $N(t)$ (because a composite action is initialized with all $h$-values being $1$). A small initial $N_R(t)$ value means that the probability of the new action to be deleted, as defined by Eq.~\ref{ProbDelet}, is high.
In order to compensate for the fast deletion of new actions we assign to each action an \textit{immunity time} of $10$ updates~\cite{simon2015evaluation}. That is, a newly created action cannot be deleted unless $10$ more rewards have been given to any sequence of actions.
Hence, we can estimate the total number of actions to be equal to $K(t)=K(0)+10+K_l(t)$, where $K_l(t)$ is the number of elements that were learned.

We used the same action composition and deletion mechanisms for automated random search. The number of actions $K_l(t)$ learned by automated random search with action composition is however always equal to zero, because automated random search does not have any other learning component.

\subsection{Details of learning performance}
Here we complement the results shown in Fig.~2(a) with an additional analysis of the agent's performance. Fig.~\ref{fig:Methods}(a) verifies that the number of experiments the PS agent has to generate before finding another $(3,3,2)$ state, continuously decreases with the number of $(3,3,2)$ states that have already been found. In order to find a $(3,3,2)$ state for the very first time the agent spends on average around $350$ experiments. This number decreases each time the agent finds a $(3,3,2)$ state. Moreover, in Fig.~\ref{fig:Methods}(a) one can observe that the length of experiments decreases with the number of $(3,3,2)$ states already found: the PS agent keeps finding on average shorter and shorter implementations that produce $(3,3,2)$ states.

Fig.~2(b) demonstrates that the PS agent finds many more interesting experiments than random search by counting the number of these experiments after $1.2\times 10^5$ time steps. Here we show that the rate at which interesting experiments are found grows over time. The fraction of interesting experiments at different times is shown in Fig.~\ref{fig:Methods}(b)-(c). The fraction of interesting experiments is calculated as the ratio of the number of new, interesting experiments to the number of all experiments performed at time step $t$. Initially, the average fraction of interesting experiments is the same for all agents and automated random search. But, in contrast to automated random search (red), the rate at which new experiments are discovered by the PS agents (blue) keeps increasing for a much longer time (Fig.~\ref{fig:Methods}(b)-(c)). Automated random search, as expected, always obtains approximately the same fraction of interesting experiments independent of the number of experiments that have already been simulated. In fact, the fraction of interesting experiments found by automated random search slowly decreases since it has the chance to encounter interesting experiments that have already been encountered. One can see that all $4$ red curves converge to certain values before they start decreasing again, indicating that on average for each automated random search agent there exists a maximum rate with which new useful experiments can be obtained.

A tantalizing assignment, related to the task of finding many interesting experiments, is that of finding many different SRVs. Even though the tasks assignment was not to find many different SRVs, this and the actual task seem to be related since PS actually found on average more different SRVs than automated random search (see Fig.~\ref{fig:Methods}(d)-(e)).

\begin{figure}[ht!]
	\begin{center}
		\includegraphics[width=8.6cm]{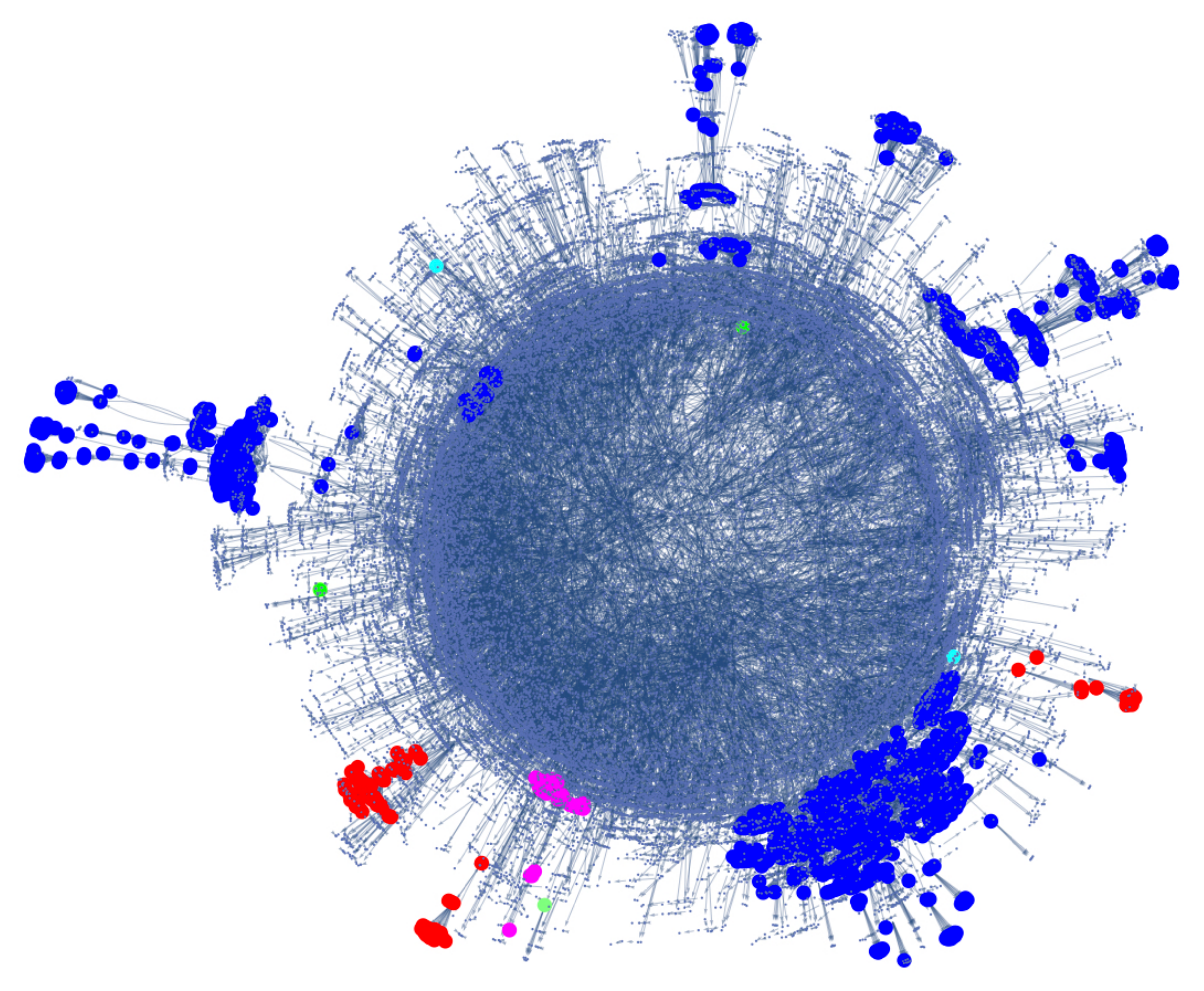}
	\end{center}
	\vspace{-0.6cm}
	\caption{Exploration space of optical setups generated by the PS agent. A maze as in Fig.~3(a), but more detailed. It was generated by our learning algorithm which has been restricted to a standard toolbox of $30$ elements as in the tasks. Each experiment consisted of at most $6$ optical elements. It can be seen that the PS exploits the clustering of interesting experiments (i.e. large colored vertices). If two nodes share a color, they have at least one SRV in common (w.r.t. a coincidence detection). Running for $1.6\times 10^4$ experiments, the graph that is shown has $34155$ nodes, about a quarter less than the randomly generated maze in Fig.~3(a).}
	\label{fig:beautifulMazePS}
\end{figure}

We have demonstrated that the PS agent is able to explore a variety of photonic quantum experiments. This suggest that there are correlations (or structural connections) between different interesting experiments which the agent exploits.
If this were not the case, we would expect to see no advantage in using PS in comparison to random search. In order to verify the existence of such correlations between different experiments we simulate the exploration space as seen by the PS agent. Compared to the exploration space as seen from the perspective of an extensive random search process shown in Fig.~3(a), the exploration space generated by the PS agent shown in Fig.~\ref{fig:beautifulMazePS} shows a more refined clustering of interesting experiments. This indicates that interesting experiments as defined here have a tendency to cluster.

\subsection{Basic optical elements}
Here we describe the standard set of optical elements $a_j$ with $j \in [1, \dots, K(0)]$ that was used to build photonic quantum experiments. There are $4$ different types of basic elements in our toolbox, which are shown in Table~\ref{tab:toolbox}: beam splitters, Dove prisms, holograms and mirrors. Non-polarizing symmetric $50/50$ beam splitters are denoted as $\mathrm{BS}_{ab}$, where indices $a$ and $b$ are the input arms of the beam splitter. There are $4$ different arms, $a$, $b$, $c$, $d$, that can be used as inputs for optical elements. Hence, there are $6$ different beam splitters that are in the basic set of optical elements. Holograms are denoted as $\mathrm{Holo}_{x,k}$ and can be placed in one arm $x$ of the four arms. $k\in\{\pm 1, \pm 2\}$ stands for a shift in the OAM of the photon. Dove prisms (DP) and mirrors (Refl) also act on single modes and can be placed in one of four arms. This gives in total $K(0) = 30$ basic optical elements, which are used throughout the paper. The action of all elements on the OAM of photons is summarized in Table~\ref{tab:toolbox}.

In addition to the described elements, a post-selection procedure is used at the end of each experiment. The desired output is conditioned on having exactly one photon in each arm. A photon in the arm $a$ is measured in the OAM basis, which leads to different possible three-photon states. Each experiment is virtually repeated until all possible three-photon states are observed and the analyzer can report whether some interesting state was observed or not.

\begin{table}[h!]
	\begin{center}
\caption{The four types of optical elements that were used in the paper and their actions on a quantum state. $m$ is the  OAM quantum number of a photon.}
\label{tab:toolbox}
\begin{tabular}{| p{3.5cm} | p{4.8cm} |}
    \hline
    Optical element & Unitary transformation \\ \hline
    $\mathrm{BS}_{ab}$: non-polarizing symmetric $50/50$ beam splitter &  $\ket{m}_a\rightarrow \left(i\ket{-m}_a + \ket{m}_b\right)/\sqrt{2}~~~~~~~$
    $\ket{m}_b\rightarrow \left(\ket{m}_a + i\ket{-m}_b\right)/\sqrt{2}$ \\[0.1cm] \hline
    $\mathrm{Holo}_{a,k}$: hologram & $\ket{m}_a\rightarrow\ket{m+k}_a$  \\[0.1cm]\hline
    $\mathrm{DP}_a$: Dove prism & $\ket{m}_a\rightarrow i\mathrm{e}^{i\pi m}\ket{-m}_a$  \\[0.1cm] \hline
    $\mathrm{Refl}_a$: mirror & $\ket{m}_a\rightarrow \ket{-m}_a$  \\[0.1cm] \hline
\end{tabular}
\end{center}
\end{table}

\subsection{Parameters of the PS agents}
For the results shown in Fig.~2(a) and Fig.~\ref{fig:Methods}(a) we used a PS agent with $\eta=1/16$, $\gamma=0$ and $\lambda=1$.
For the results shown in Fig.~2(b) and Fig.~\ref{fig:Methods}(b)--(e) we used a PS agent (solid blue) with $\eta=0.1$, $\gamma=10^{-3}$ and $\lambda=100$ for all values of $L$. The PS agent with action composition (dashed blue) had the following parameters: $\eta=0.1$, $\gamma=5\times10^{-4}$, $\lambda=50$ for $L=6, 8$ and $\lambda=25$ for $L=10, 12$.

\end{document}